\documentclass[12pt]{article}
\usepackage{graphicx}
\usepackage{epsfig}
\begin{document}

\begin{center}
{\bf HIGH ENERGY ELASTIC pp SCATTERING
IN ADDITIVE QUARK MODEL}

\vspace{.2cm}

Yu.M. Shabelski and A.G. Shuvaev \\

\vspace{.5cm}

Petersburg Nuclear Physics Institute, Kurchatov National
Research Centre\\
Gatchina, St. Petersburg 188300, Russia\\
\vskip 0.9 truecm
E-mail: shabelsk@thd.pnpi.spb.ru\\
E-mail: shuvaev@thd.pnpi.spb.ru

\vspace{1.2cm}

\end{center}

\begin{abstract}
\noindent
High energy $pp$ and $p\bar p$ elastic scattering is treated
in the framework of Additive Quark Model. The reasonable
agreement with experimental data is achieved with the natural
parameters for the strong matter distribution inside proton.
\end{abstract}


\section{Introduction}
Regge theory provides a useful tool for the phenomenological description
of high energy hadron collisions~\cite{Dr, RMK, Khoze:2013dha,
Gotsman:2012rm, Gotsman:2012rq, MerShab, Sel}.
The quantitative predictions of Regge calculus are essentially dependent
on the assumed coupling of participating hadrons to Pomeron. In this
paper high energy elastic $pp$ and $p\bar p$ scattering data including the recent
LHC ones are analyzed in terms of the simple Regge exchange approach
in the framework of Additive Quark Model (AQM)~\cite{LF, VH}.
In AQM baryon is treated as a system of
three spatially separated compact objects -- constituent quarks.
Each constituent quark is colored, has internal quark-gluon
structure and finite radius that is much less than the radius of
proton, $r_q \ll r_p$. This picture is in good agreement both with
$SU(3)$ symmetry of strong interaction and the quark-gluon structure
of proton~\cite{DDT, Shekhter}.

The three constituent quarks are assumed in AQM to be the incident
particles in $pp$ or $p\bar p$ scattering.
Elastic amplitudes for the large energy $s=(p_1+p_2)^2$
and the small momentum transfer $t$ are dominated by the Pomeron
exchange. We neglect the small difference in $pp$ and $p\bar p$
scattering coming from the exchange of negative signature Reggeons,
Odderon (see e.g.~\cite{Avila} and references therein), $\omega$-Reggeon
etc, since their contribution is suppressed by $s$.

The single $t$-channel exchange results into $s$-channel amplitude
of the constituent quarks scattering
\begin{equation}
\label{Mqq}
M_{qq}^{(1)}(s,t) = \gamma_{qq}(t) \cdot
\left(\frac{s}{s_0}\right)^{\alpha_P(t) - 1}
\cdot \eta_P \;,
\end{equation}
where $\alpha_P(t) = \alpha_P(0) + \alpha^\prime_P\cdot t$ is
the Pomeron trajectory specified by the values of intercept,
$\alpha_P(0)$, and slope, $\alpha^\prime_P$.
The (positive) signature factor,
$$
\eta_P \,=\, i \,-\, \tan^{-1}
\left(\frac{\pi \alpha_P(t)}2\right),
$$
determines the complex structure of the amplitude. The factor
$\gamma_{qq}(t)=g_1(t)\cdot g_2(t)$ has the meaning
of the Pomeron coupling to the beam and target particles,
the functions $g_{1,2}(t)$ being the vertices of constituent
quark-Pomeron interaction (filled circles in Fig.~\ref{1P2P}).
As the same set of the Pomeron parameters $\Delta=\alpha_P(0)-1$,
$\alpha^\prime_P$ and $\gamma_{qq}$ describes proton
and antiproton scattering, both $pp$ and $p\bar p$
data have been commonly used to fix their values
\footnote{Strictly speaking, exchange of the positive signature
Reggeons determine half of the sum of $pp$ and $p\bar p$ elastic
amplitude. We neglect their difference.}.

The one-Pomeron exchange between
two protons includes the sum over all possible exchanges between
the quark pairs~\cite{LF,VH}. Each term in the sum has
a form~(\ref{Mqq}) with the functions $g_{1,2}(t)$ attributed
to the individual constituent quarks.

Due to factorization of longitudinal and transverse degrees
of freedom the longitudinal momenta are integrated over separately
in high energy limit. After this the transverse part of the quark
distribution is actually relevant only.
Writing the proton wavefunction in the transverse momentum space
as $\psi(k_1,k_2,k_3)$, where $k_i$ are the quark transverse momenta,
\begin{equation}
\label{norm}
\int \bigl|\psi(k_1,k_2,k_3)\bigr|^2\delta^{(2)}(k_1+k_2+k_3)\,
d^2k_1 d^2k_2 d^2k_3\,=\,1,
\end{equation}
the proton - Pomeron vertex, $F_P(Q,0,0)$, $t=-Q^2$,
is given by the overlap function
\begin{eqnarray}
\label{FP}
F_P(Q_1,Q_2,Q_3)\,&=&\,\int \psi^*(k_1,k_2,k_3)\,
\psi(k_1+Q_1,k_2+Q_2,k_3+Q_3) \\
&&\times\,\delta^{(2)}(k_1+k_2+k_3)\,d^2k_1 d^2k_2 d^2k_3.
\nonumber
\end{eqnarray}
The function $F_P(Q,0,0)$ plays a role of the proton formfactor
for strong interaction in AQM (see section~2).

In what follows we assume the Pomeron trajectory in the simplest form
$$
\left(\frac{s}{s_0}\right)^{\alpha_P(t) - 1}\,=\,e^{\Delta\cdot\xi}
e^{-r_q^2\,q^2}, ~~ \xi\equiv \ln\frac{s}{s_0},~~
r_q^2\equiv \alpha^\prime\cdot\xi.
$$
The value $r_q^2$ defines the radius of quark-quark interaction
while $S_0=(9~{\rm GeV})^2$ has the meaning of typical energy scale
in Regge theory. Putting together all 9 equal quark-quark contributions
(one of them is shown in Fig.~\ref{1P2P}~a) we get the first order
elastic $pp$ (or $p\bar p$, we do not distinguish between them here)
amplitude
\begin{equation}
\label{M1}
M_{pp}^{(1)}\,=\,9\biggl(\gamma_{qq}\eta_P e^{\Delta\cdot\xi}\biggr)\,
e^{-r_q^2\,Q^2}F_P(Q,0,0)^2.
\end{equation}

Actually the formula (\ref{M1}) with a single Pomeron
gives amplitude in the impulse approximation~\cite{LF}.
Similarly to light nuclear scattering the multipomeron exchange
should be added (see Glauber theory \cite{Glaub, FG}) giving rise
to the terms $M_{pp}^{(2)}$, $M_{pp}^{(3)}$ etc, so that the total
amplitude
$$
M_{pp}\,=\,\sum_n M_{pp}^{(n)}.
$$
If $r_q/r_p \to 0$ the multiple interactions become negligible
leaving in the sum the first term only.

The optical theorem, that relates the total elastic cross section
and imaginary part of the amplitude, in the normalization adopted
here reads
$$
\sigma_{pp}^{tot}\,=\,8\pi\,{\rm Im}\, M_{pp}(s,t=0).
$$
The differential cross section is evaluated in this normalization as
\begin{equation}
\label{ds/dt}
\frac{d\sigma}{dt}\,=\,4\pi\,\bigl|M_{pp}(s,t)\bigr|^2\,
=\,4\pi\,\bigl[\bigl({\rm Re}\, M_{pp}(s,t)\bigr)^2
+ \bigl({\rm Im}\, M_{pp}(s,t)\bigr)^2\bigr].
\end{equation}

Interference of the contributions generated by the various number
of Po\-me\-r\-ons leads to the occurrence of local minima in the differential
elastic cross section. Experimentally the minimum at $t\simeq 0.53$~GeV$^2$
for the energy $\sqrt s = 7$~TeV is well observed at LHC \cite{TO1,TO2}.
The minima at another $t$ are also possible. Basically there is interplay
between the minima in real and imaginary parts of the amplitude in
the expression (\ref{ds/dt}) so that the minimum in the imaginary part
could be filled by the large real part at the same $t$.

The present paper aims to give theoretical description
of the experimental $pp$ and $p\bar p$ elastic scattering data
in the energy interval 546~Gev $\div$ 7~Tev in the framework of AQM.
The analysis of the differences in $pp$ and $p\bar p$ amplitudes
needs a more subtle treatment than that involving only Pomeron
exchanges and goes beyond our approach.

\section{Elastic Scattering Amplitude in AQM}

\begin{figure}[htb]
\centering
\includegraphics[width=0.6\hsize]{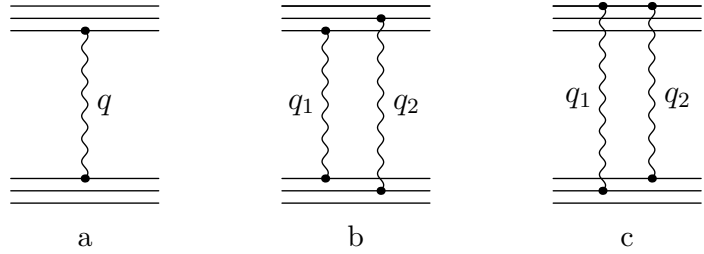}
\caption{\footnotesize The AQM diagrams for $pp$ elastic scattering. The
straight lines stand for quarks, the waved lines denote Pomerons, $Q$ is the
momentum transferred, $t=-Q^2$. Diagram~(a) is the one of the single Pomeron
diagrams, diagrams~(b) and~(c) represent double Pomeron exchange with two
Pomeron coupled to the different quark~(b) and to the same quarks~(c),
$q_1+q_2=Q$.}
\label{1P2P}
\end{figure}

\begin{figure}[htb]
\centering
\includegraphics[width=.8\hsize]{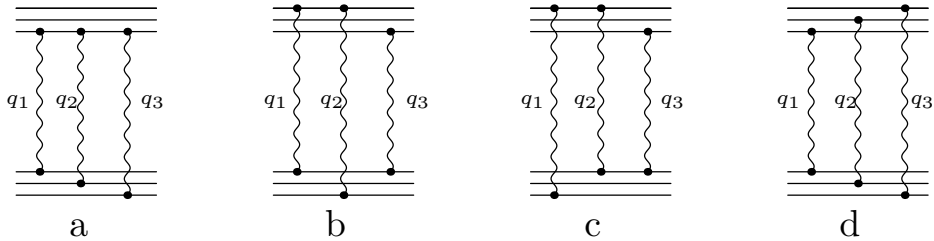}
\caption{\footnotesize
Several AQM diagrams of the third order.}
\label{3P}
\end{figure}

In AQM there are a total 9 orders of interactions.
The first order is the sum of all interactions between single
$qq$ pairs \footnote{Note that one $qq$ pair interaction
in AQM may include the contributions of several Gribovs' Pomerons
\cite{Gribov}. We assume that high energy $qq$ scattering is
described by single effective Pomeron exchange between each
$qq$ pair, the parameters of this effective Pomeron could be different
from those of Gribov's bare Pomerons.} It contains 9 terms. Similar to Glauber
theory \cite{Glaub, FG} one has to rule out
the multiple interactions between the same quark pair.
AQM permits the Pomeron to connect any two quark lines only once.
It crucially decreases the combinatorics leaving the diagrams
with no more than 9 effective Pomerons. Examples of various order
diagrams are shown in Fig~\ref{1P2P},\ref{3P}.

Let $q_i$ be the transverse momentum carried by the $i$-th
effective Pomeron, $Q_k$ and $Q_l^\prime$ denote the momenta
transferred to the quark $k$ from the target proton or quark
$l$ from the beam proton during the scattering process.
If no Pomerons are attached to the quark $j$, that is it does not
interact, then $Q_j=0$. If only one Pomeron carrying momentum $q_i$
is attached to the quark $j$, then $Q_j = q_i$. If two Pomerons
with the momenta $q_i$ and $q_k$ are coupled to it, then $Q_j = q_i+q_k$.
In other words,
$$
Q_k\,=\,\sum q_i~~{\rm if~Pomeron}~i~{\rm is~attached~to~quark}~k,
$$
$$
Q_l^\prime\,=\,\sum q_{i^\prime}~~{\rm if~Pomeron}~i^\prime~
{\rm is~attached~to~quark}~l.
$$
With these notations the $n$ order amplitude is equal to
\begin{eqnarray}
M^{(n)}\,&=&\,i^{n-1}\biggl(\gamma_{qq}\eta_P e^{\Delta\cdot\xi}\biggr)^n\,
\int\frac{d^2q_1}{\pi}\cdots \frac{d^2q_n}{\pi}
\,\pi\,\delta^{(2)}(q_1+\ldots +\,q_n-Q)\,
\nonumber \\
\label{Mn}
&&\times\,e^{-r_q^2(q_1^2+\ldots + q_n^2)}\,
\frac 1{n!}\sum\limits_{n~\rm connections}
F_P(Q_1,Q_2,Q_3)\,F_P(Q_1^{\,\prime},Q_2^{\,\prime},Q_3^{\,\prime}),
\end{eqnarray}
where the sum in the last factor is taken over all distinct ways
to connect the pairs of beam and target quarks by $n$ effective
Pomerons, each pair being connected no more than once.
The permutations of identical Pomerons in the integrals is compensated
by $1/n!$ in front the sum.

There are two types of diagrams in the second order,
\begin{eqnarray}
&&\frac 1{2!}\sum\limits_{2~\rm connections}
F_P(Q_1,Q_2,Q_3)\,F_P(Q_1^{\,\prime},Q_2^{\,\prime},Q_3^{\,\prime})
\nonumber \\
&&\,=\,
18\,F_P(q_1,q_2,0)F_P(q_1,q_2,0)\,+\,18\,F_P(Q,0,0)F_P(q_1,q_2,0),
~~Q\,=\,q_1\,+\,q_2,
\nonumber
\end{eqnarray}
where the first term comes from the diagrams with both Pomerons
coupled to the different quark lines whereas in the second one they
are attached to the same line (Fig.~\ref{1P2P}~b,c).

The third order sum is
\begin{eqnarray}
&&\frac 1{3!}\sum\limits_{3~\rm connections}
F_P(Q_1,Q_2,Q_3)\,F_P(Q_1^{\,\prime},Q_2^{\,\prime},Q_3^{\,\prime})
\nonumber \\
\,&=&\,
6\,F_P(Q,0,0)F_P(q_1,q_2,q_3)
+ 9\,F_P(q_1 + q_2,0,q_3)F_P(q_1 + q_3,0,q_2) \nonumber \\
&&+ 9\,F_P(q_1 + q_2,0,q_3)F_P(q_2 + q_3,0,q_1)
+ 6\,F_P(q_1 + q_2,0,q_3)F_P(q_1,q_3,q_2)  \nonumber \\
&&+ 18\,F_P(q_1 + q_3,0,q_2)F_P(q_2 + q_3,0,q_1)
+ 15\,F_P(q_1 + q_3,0,q_2)F_P(q_1,q_3,q_2)\nonumber \\
&&+ 15\,F_P(q_2 + q_3,0,q_1)F_P(q_1,q_3,q_2)
+ 6\,F_P(q_3,q_2,q_1)F_P(q_1,q_2,q_3), \nonumber
\end{eqnarray}
$$
Q\,=\,q_1\,+\,q_2\,+\,q_3.
$$
Here the first term arises when all three Pomerons are attached
to the same line, in the last term they connect three different
quarks, diagrams (a) and (d) in the Fig.~\ref{3P}. The second
and the third terms correspond to the diagrams (b) and (c). The rest
terms are provided by various permutations of the Pomeron lines
in these diagrams, in particular, by flipping a diagram as a whole.
The numerical coefficients encounter the number of connections
resulting into equal expressions after variables changing in
the integrals~(\ref{Mn}).

In the highest order, containing 9 effective Pomerons,
$$
\frac 1{9!}\sum\limits_{9~\rm connections}
F_P(Q_1,Q_2,Q_3)\,F_P(Q_1^\prime,Q_2^\prime,Q_3^\prime)
$$
$$
=\,F_P(q_1 + q_2 + q_3,q_4 + q_5 + q_6,q_7 + q_8 + q_9)
F_P(q_1 + q_4 + q_7,q_2 + q_5 + q_8,q_3 + q_6 + q_9),
$$
each quark from the beam proton interacts with the quark from
the target one. The other orders, $3<n<9$, have similar
but more lengthly structure due to large combinatorics
to redistribute $q_1,\ldots, q_n$ momenta among $Q_i$
and $Q_i^\prime$ groups~\footnote{The full set of similar diagrams
for $pp$ and $\alpha\alpha$ scattering can be found
in (\cite{Bialas:1977xp, Bialas:1982rd}).}.

The function $F_P(q_1,q_2,q_3)$ is determined by the proton
wavefunction in terms of the constituent quarks (\ref{FP}).
We model the transverse part of the wavefunction in a simple
form of two gaussian packets,
\begin{equation}
\label{gausspack}
\psi(k_1,k_2,k_3)\,=\,N\bigl[\,e^{-a_1(k_1^2+k_2^2+k_3^2)}\,
+\,C\,e^{-a_2(k_1^2+k_2^2+k_3^2)}\bigr],
\end{equation}
normalized to unity (\ref{norm}). One packet parametrization,
$C=0$, is insufficient to reproduce experimental data
as imposing too strong mutual dependence between the total
cross section, the minimum position and the value of the slope
at $t=0$.

All parameters used in the calculation naturally fall into
two different kinds: the parameters of the Pomeron and those
specifying the structure of colliding particles.
The former type, $\Delta$, $\alpha^\prime$, $\gamma_{qq}$,
refers to the high energy scattering theory while the latter,
$a_{1,2}$ and $C$, details the matter distribution inside
the proton in the low energy limit(similar to density distribution
in atomic nuclear).

\begin{figure}[htb]
\centering
\includegraphics[width=.4\hsize]{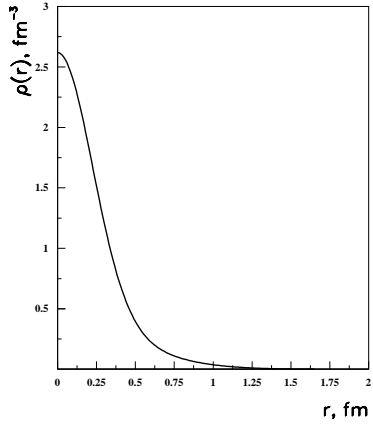}
\vskip -1.5 cm
\caption{\footnotesize
The radial density distribution in the proton
calculated with the wavefunction
(\ref{gausspack}) and parameters (\ref{set}).}
\label{rho}
\end{figure}

With the chosen values of $a_{1,2}$ and $C$
(see below) the wavefunction (\ref{gausspack}) results into
the density shown in Fig.~\ref{rho}
It looks rather naturally
having mean squared radius $\sqrt{<r^2>}=0.68$~fm, that is
close to the electromagnetic radius.
Some discrepancy can be explained by the difference
in the distributions of charged and strong interacting matter
inside the proton.

\section{Comparison with the experimental data}

Here we present the numerical results obtained by summing over
all 9 orders of interaction of the Additive Quark Model for
the energies in the interval $\sqrt s=546$~GeV $\div$ 7~TeV.
They are compared with the experimental data taken
from~\cite{TO1, TO2, Amos:1985wx, Abe:1993xx, Bozzo:1985th,
Kwa, Am, Amos:1990fw}.
The following set of parameters have been used
\begin{equation}
\label{set}
\begin{array}{lll}
\Delta=0.107~&
\alpha^\prime=0.32\,{\rm GeV}^{-2}~&
\gamma_{qq}=0.44\,{\rm GeV}^{-2}~\\
a_1=4.8\,{\rm GeV}^{-2}~&
a_2=0.87\,{\rm GeV}^{-2}~&
C=0.132
\end{array}.
\end{equation}
In choosing these values we have not distinguished
$pp$ and $p\bar p$ data.

\begin{figure}[htb]
\vskip -1.cm
\includegraphics[width=.45\textwidth]{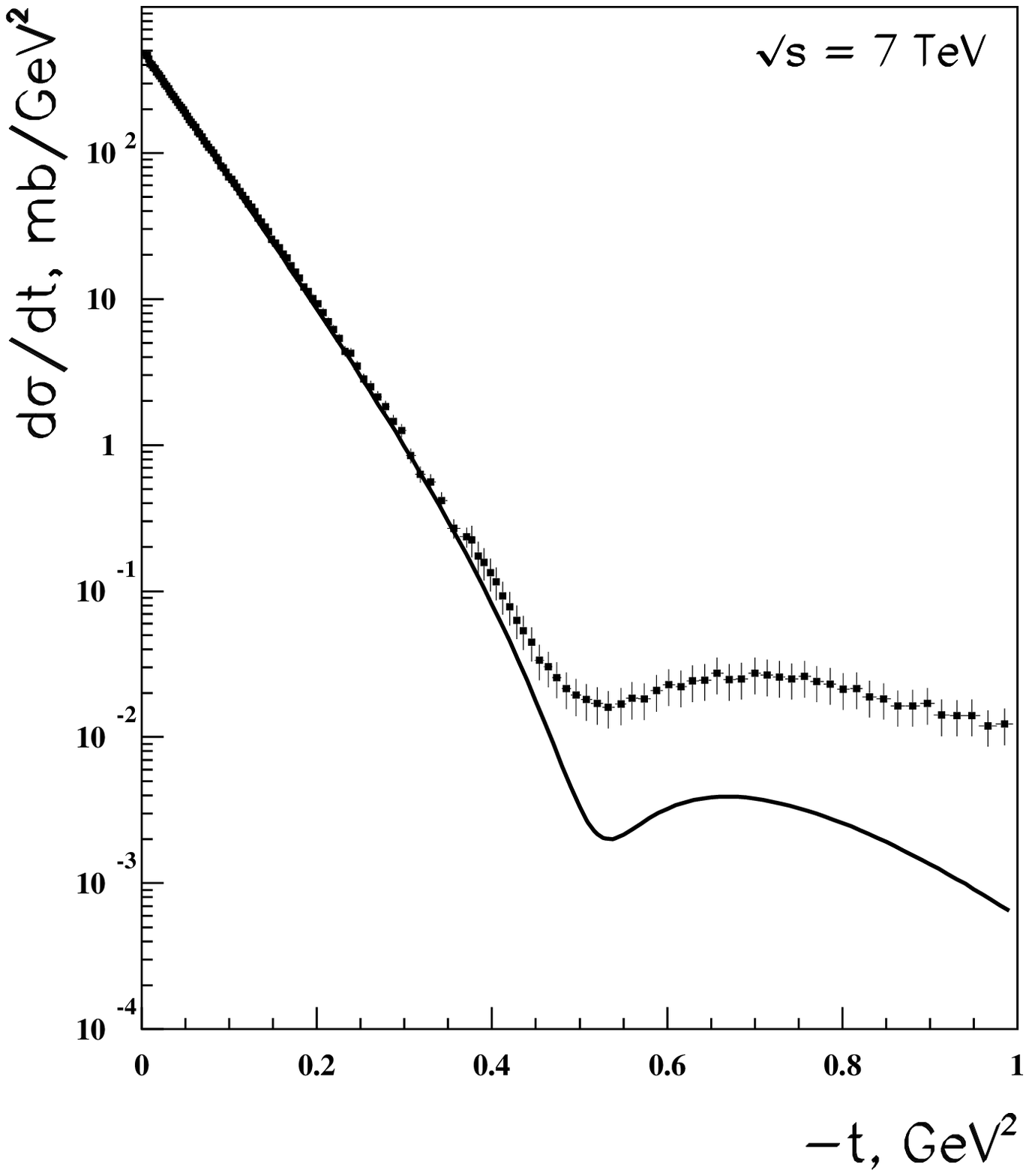}
\includegraphics[width=.45\textwidth]{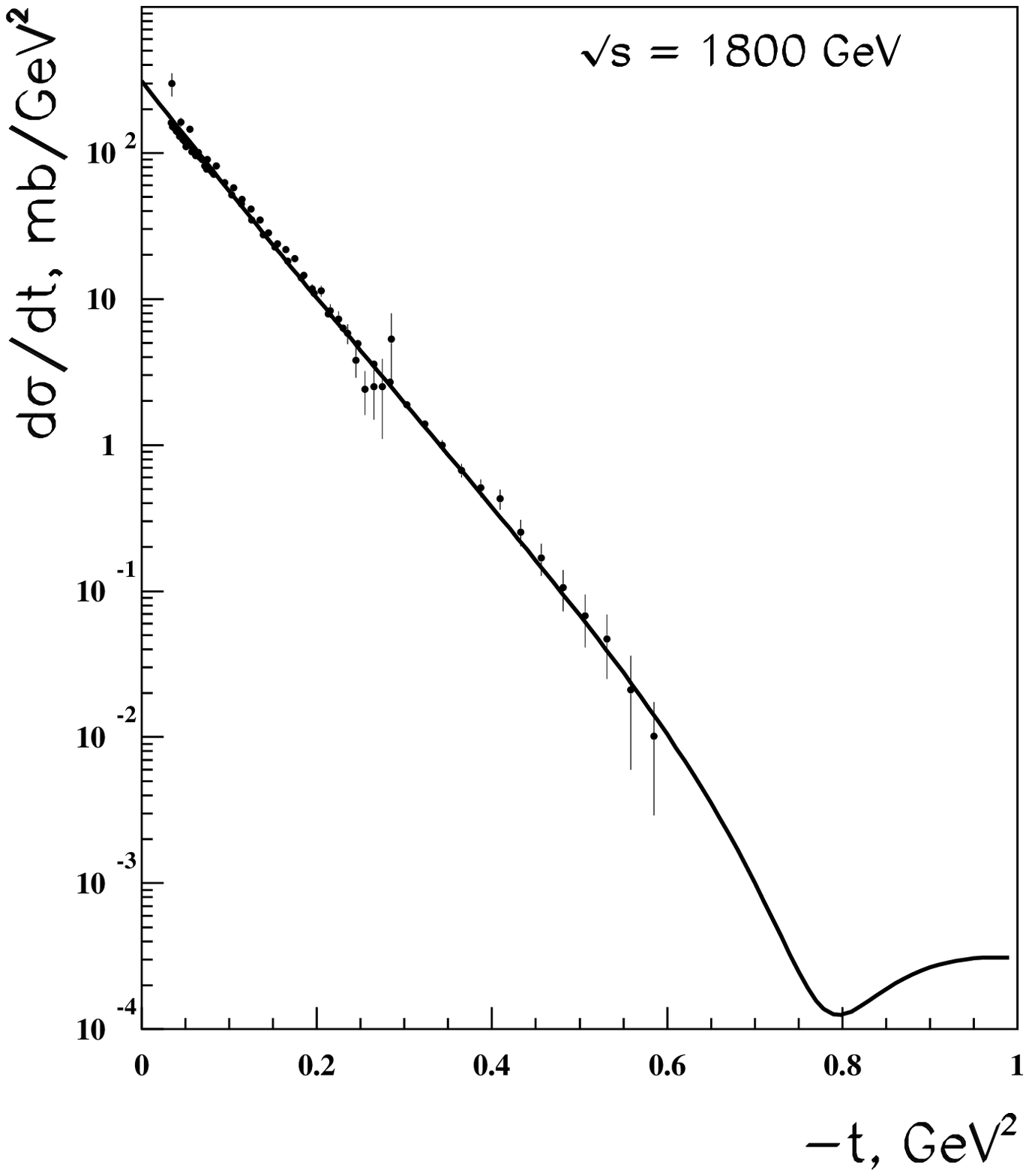}
\vskip -1.5 cm
\caption{\footnotesize
The differential cross section of elastic $pp$ scattering
at $\sqrt s = 7$~TeV and $p\bar p$ scattering at
$\sqrt s = 1800$~GeV.
The experimental points have been taken from
\cite{Abe:1993xx,
Amos:1990fw,TOT1a,TO1, TO2}.}
\label{1800}
\end{figure}
\begin{figure}[htb]
\vskip -1.cm
\includegraphics[width=.45\textwidth]{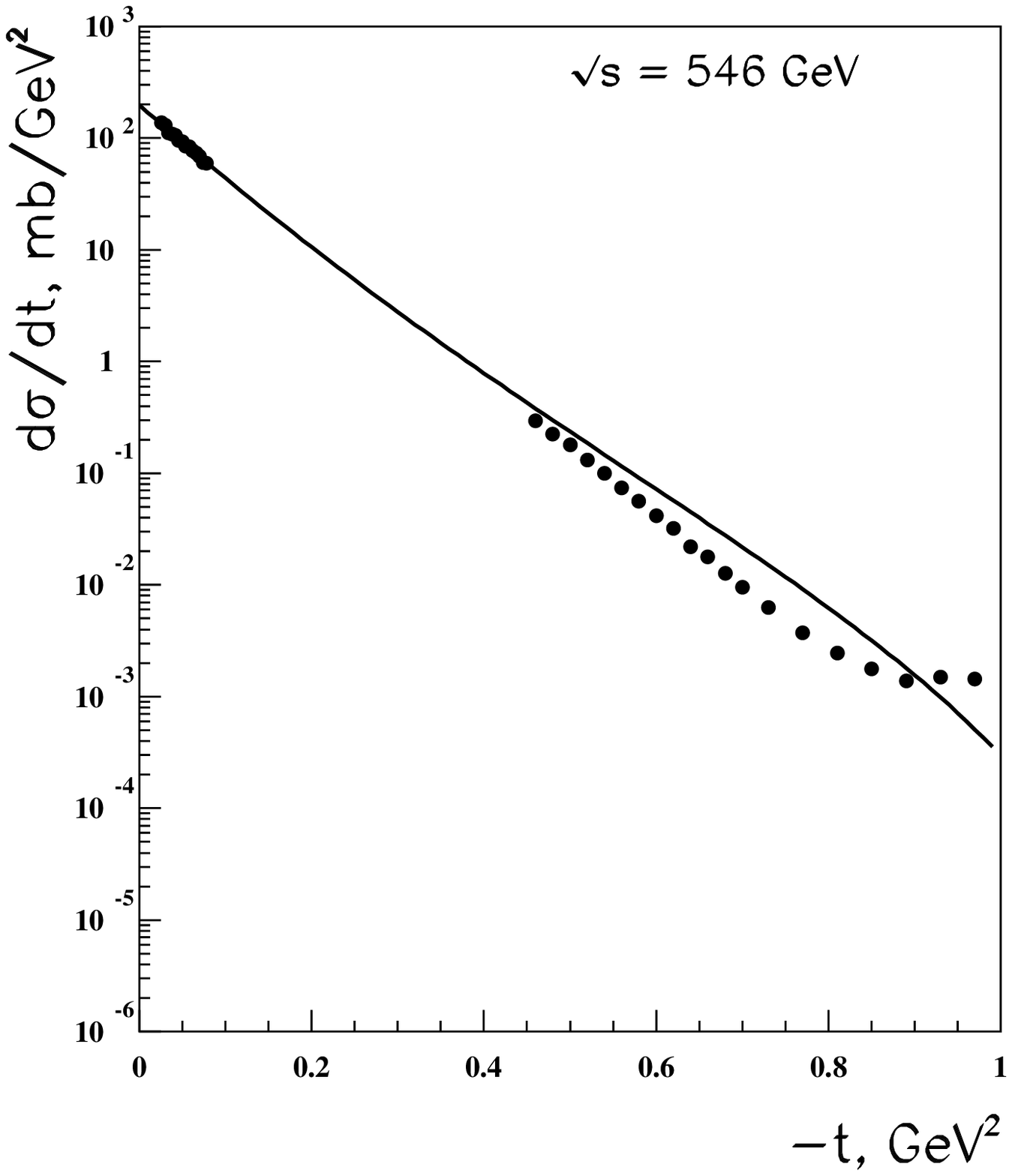}
\includegraphics[width=.45\textwidth]{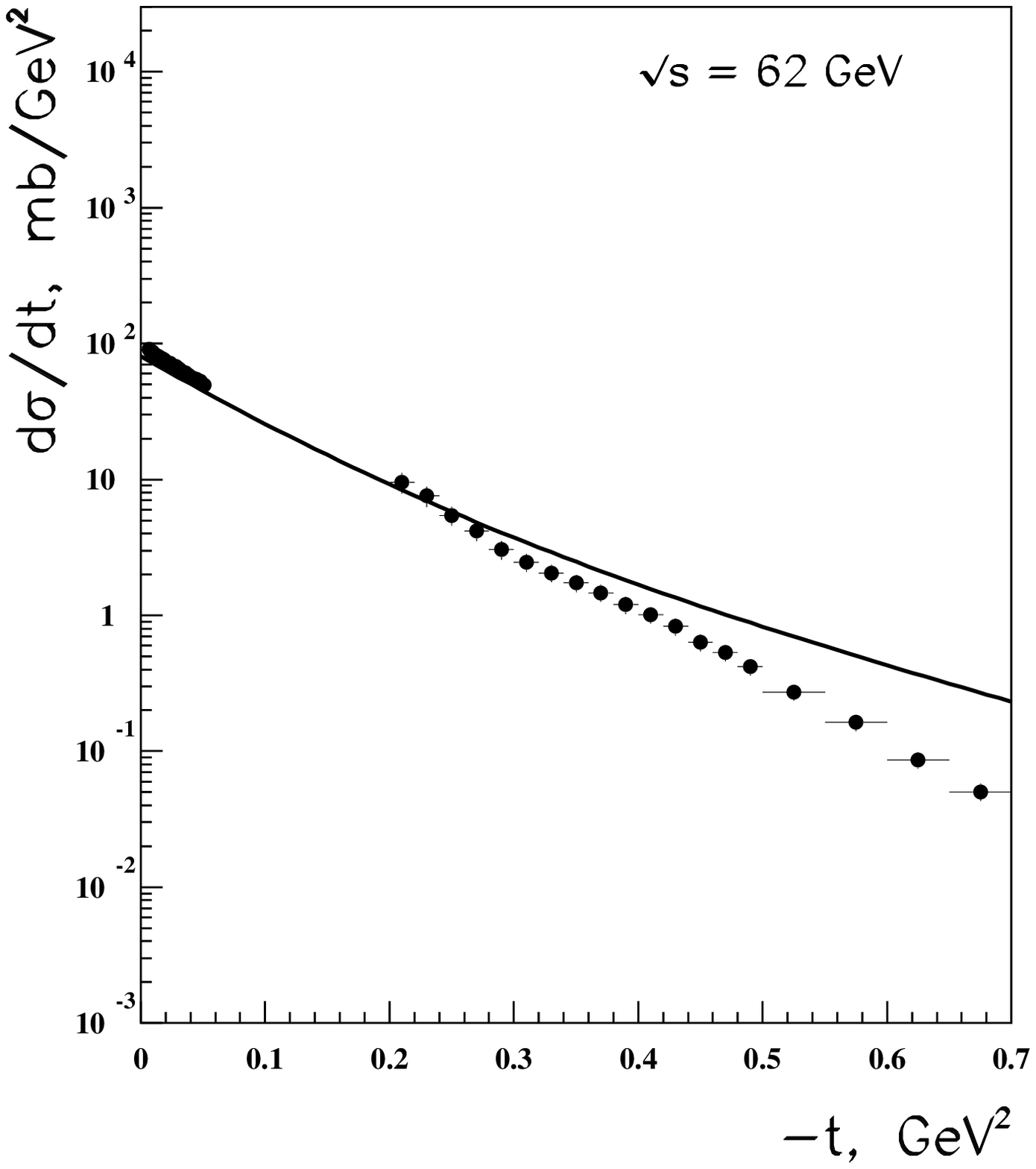}
\vskip -1.5 cm
\caption{\footnotesize
The differential cross section of elastic $p\bar p$ scattering
at $\sqrt s = 546$~GeV and elastic $p\bar p$ and $pp$ scattering
at $\sqrt s = 62$~GeV.
The experimental points have been taken from
\cite{Abe:1993xx, Bozzo:1985th} (left panel),
\cite{Amos:1985wx} for $p\bar p$ and \cite{Kwa} for $pp$
(right panel).}
\label{62}
\end{figure}

The calculated values of the total cross section $\sigma^{tot}$, of
$d\sigma/dt(t=0)$, of the slope of the elastic cross section $B$ ($d\sigma/dt
\sim \exp(-B\cdot t))$, and of the ratio $\rho={\rm Re}\,\sigma/{\rm
Im}\,\sigma(t=0)$ are presented in the Table~1 along with the experimental
data. The theoretical slope is calculated within the interval $|\,t\,|=0 -
0.1$~GeV$^2$. Here we again neglect the difference between $pp$ and $p\bar p$
collisions in the presented data.
\begin{center}
\begin{tabular}{|c||c|c|c|c|} \hline
$\sqrt{s}$  & $\sigma^{tot}$ (mb) & $d\sigma/dt(t=0)$
& $B$ (GeV$^{-2}$) & ${\rm Re}\,\sigma/{\rm Im}\,\sigma$ \\
& & (mb/GeV$^2$)& &$(t=0)$ \\ \hline
7 TeV & 98.54 & 500.32 & 20.16 & 0.099 \\
\cite{TOT1a} & $98.3 \pm 2.8$ & - & $20.1 \pm 0,4$ & -  \\ \hline
1.8 TeV   & 77.58 & 310.48 & 17.33 & 0.104 \\
\cite{Abe:1993xx} & - & $334.6 \pm 18.8$ & $16.98 \pm 0,25$ & - \\ \hline
546 GeV & 62.06 & 198.84 & 15.09 & 0.11  \\
\cite{Auger}& - & - & - & $0.135 \pm 0.015$ \\
\cite{Abe:1993xx} & - & $196.1 \pm 6$ & $15.35 \pm 0,19$ & - \\ \hline
62 Gev & 39.54 & 80.73 & 11.46 & 0.11 \\
\cite{Amos:1985wx} & $43.55 \pm 0.31$ & - & $13.02 \pm 0.27$ & - \\
\cite{Am} & - & - & $13.3 \pm 0.3$ & $0.095 \pm 0.011$\\ \hline
\end{tabular}
\end{center}
\vskip-0.2cm
\noindent
Table 1. The comparison of the calculated values of total cross sections
$\sigma^{tot}$, of $d\sigma/dt(t=0)$, slope parameter $B$
and ratio ${\rm Re}\,\sigma/{\rm Im}\,\sigma(t=0)$ with the available
experimental data.
\vskip0.2cm
The values of the total cross sections and the slopes are in reasonable
agreement with the experimental data, the value
$\rho={\rm Re}\,\sigma/{\rm Im}\,\sigma(t=0)$ is also well reproduced
at $\sqrt s = 546$~GeV.
The AQM assumption $r_q^2 \ll r_p^2$ continues to hold satisfactorily even
at the LHC energy $\sqrt s = 7$~TeV for $r_q^2 \simeq 5~{\rm GeV}^{-2}$,
$r_p^2 \simeq 12~{\rm GeV}^{-2}$. Probably it would be better fulfilled
if the Pomeron slope $\alpha^\prime$ would be lesser.
The value obtained for the total cross section for
$p\bar p$ scattering at $\sqrt s = 62$~GeV
are smaller than the experimental data. The reason might be
in non-Pomeron contributions, for example in $f$~Reggeon,
whose effect could be significant for the low energies
but disappears when the energy grows.

The experimental position of the local minimum of the differential
$pp$ cross section at $\sqrt s=7$~Tev is also reproduced although
the theory predicts the more deep minimum.
As the transferred momentum increases,
$|\,t\,| \ge 1$~GeV$^2$, the interaction becomes sensitive
to the internal structure of constituent quarks that have
a finite size.
We have used a simple parametrization for their radius
$r_q^2=\alpha^\prime \xi$, but there is a possibility
for additional constants or more complicated functions.
\begin{figure}[htb]
\centering
\vskip -1.cm
\includegraphics[width=.45\textwidth]{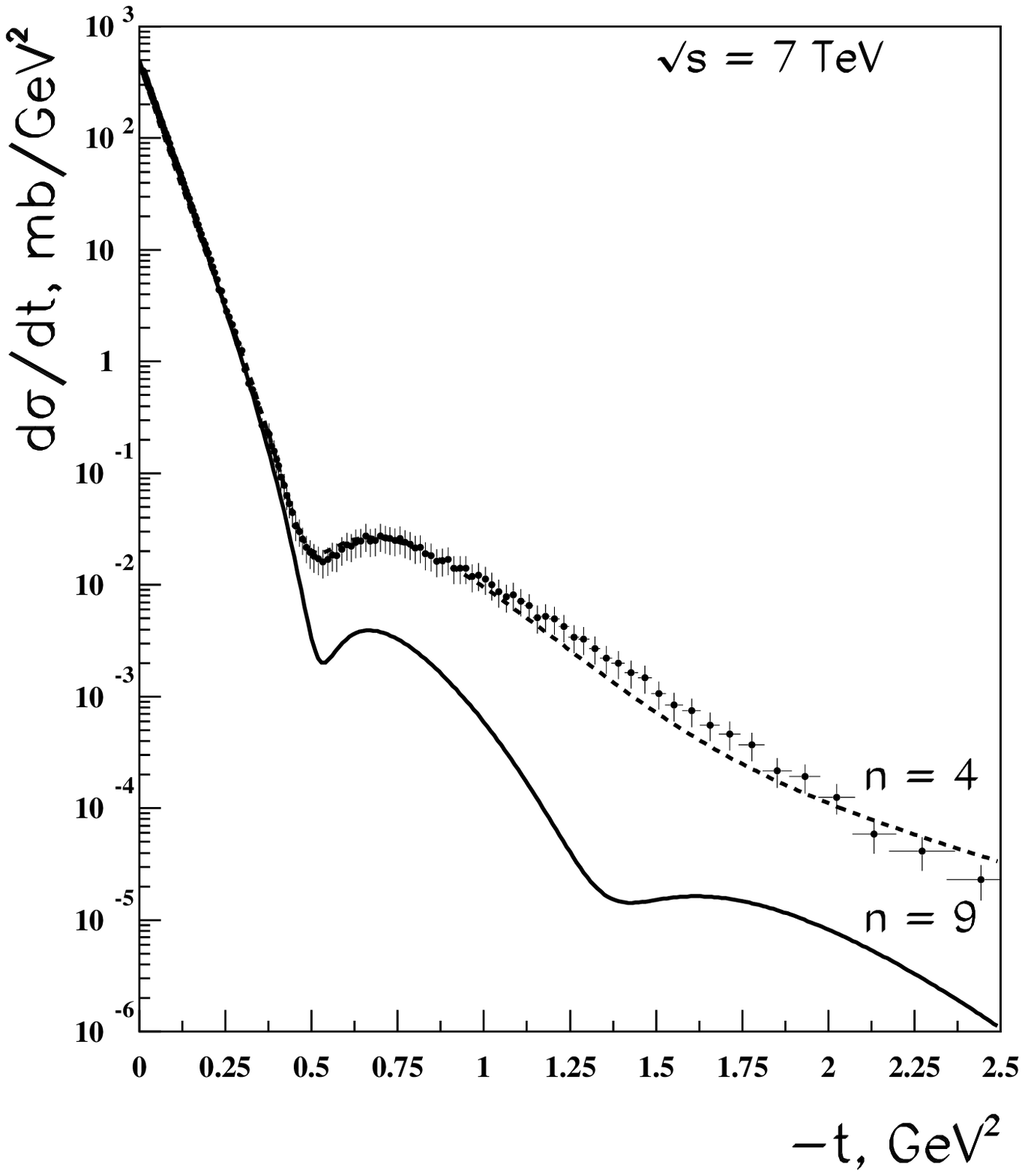}
\includegraphics[width=.45\textwidth]{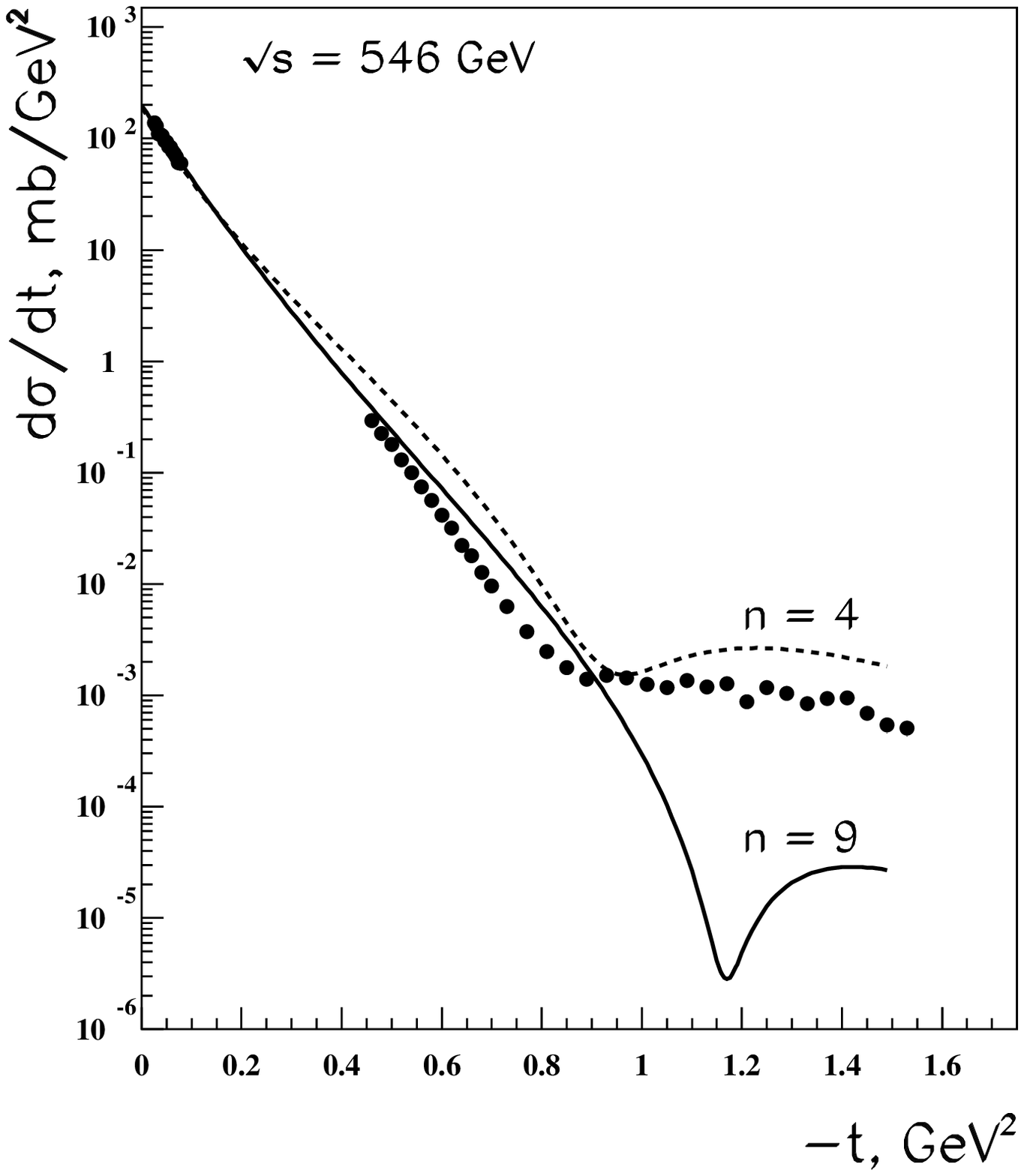}
\vskip -1.5 cm
\caption{\footnotesize
The differential cross section of elastic $pp$ scattering
at $\sqrt s = 7$~TeV (left panel) and elastic $p\bar p$
scattering at $\sqrt s = 546$~GeV (right panel).
The dashed line ($n=4$) presents the sum of first
four orders of AQM calculated with the parameters $\Delta = 0.1$,
$\alpha^\prime = 0.38$~GeV$^{-2}$, $\gamma = 0.45$~GeV$^{-2}$,
$a_1 = 6.5$~GeV$^{-2}$, $a_2 = 0.47$~GeV$^{-2}$, $C = 0.081$.
The solid line ($n=9$) presents the sum of all AQM orders for
the parameters (\ref{set}). It exhibits the second local minimum
for $pp$ scattering at $\sqrt s = 7$~TeV.
The experimental points have been taken from \cite{TOT1a,TO1, TO2,
Abe:1993xx, Bozzo:1985th}}.
\label{4-9}
\end{figure}
The experimental data for $p\bar p$ scattering show no clean dips
at the smallest energies $\sqrt s = 1800$~GeV, 546~GeV and 62~GeV.
The reason, probably, is in substantial contribution that could be
made to $p\bar p$ amplitude by by the negative signature Reggeons
(see Conclusion).

It is important to note that the complete 9 orders AQM calculation
yields a second local minimum of the differential elastic $pp$ cross
section at $\sqrt s = 7$~TeV for $|\,t\,|\ge 1$~GeV$^2$ unobservable
experimentally. It may indicate to an invalidity of our description
for $|\,t\,| > 1$~GeV$^2$, where the internal structure
of the constituent quarks becomes important.
From another point of view there is an uncontrolled contribution of
multipomeron diagrams, (e.g. enhanced Pomeron
diagrams~\cite{Gribov}) rapidly growing with the number of Pomerons.
Their numerical value is determined by unknown vertices of $m \to n$
Pomeron transitions, which increases the uncertainty of the next
orders. In this context, it is worth stressing that the sum of only
first four orders of AQM results (with a little modified parameters)
into the theoretical curve that better fits the LHC $pp$ data at
the more broad $t$ interval and does not exhibit an extra local minimum
as is shown in Fig.~\ref{4-9} left.

An example of AQM predictions for $p\bar p$ scattering in the region
$|\,t\,| \ge 1$~GeV$^2$ for the energy $\sqrt s = 546$~GeV
is presented in Fig.~\ref{4-9} right .
The theory gives here the large dip, which is in fact absent
in the experimental data. It points to the essential
effect of the contributions, that are not accounted for in AQM,
such as negative signature Reggeons etc.
Nevertheless, the first four orders of AQM better describe
the cross section behavior (give much smaller dip than the complete
sum) even in this case. Both theoretical curves, $n=4$ and $n=9$,
have no extra local minima up to $|\,t\,| \simeq 2.5$~GeV$^2$
for this energy.

\section{Conclusion}

We show that the simple AQM model gives reasonable description
of the high energy $pp$ and $p\bar p$ elastic scattering at
the not large momentum transferred, $|\,t\,| \le 1$~Gev$^2$,
that is at the distances where the internal structure
of the constituent quarks probably do not show up.
Our model contains 6 parameters, but only 3 of them, ($\Delta$,
$\alpha^\prime$, $\gamma_{qq}$), are employed to describe high energy
scattering while 3 others ($a_{1,2}$, $C$) determine the matter
distribution in the proton. Note that the first group
of parameters refers to the Pomeron and has been chosen over
together taken $pp$ and $p\bar p$ data. The interesting fact
is that the matter distribution cannot be parameterized
by one Gaussian packet (see also \cite{MerShab}).

As mentioned in the Introduction we neglect the contribution of
the negative signature Reggons that leads to the difference in $pp$ and
$p\bar p$ scattering especially around the dip.
Some difference between them is experimentally
observed \cite{Breakstone:1985pe,Erhan:1984mv} at $\sqrt s = 53$~GeV
near the dip ($|\,t\,|> 1$~GeV$^2$). The energy behavior
of this difference depends on the value of the Odderon intercept,
$\alpha_{\rm Odd}$. The same Odderon effects are responsible
for the difference in $p$ and $\bar p$ yields in the central
region of the $pp$ collision. However, it was shown
\cite{Aamodt:2010dx,Merino:2009zz} that the observed
ratio $\bar p/p$ is compatible with the value
$\alpha_{\rm Odd} \sim 0.5$. In this case the difference in
$d\sigma/dt$ for $pp$ and $p\bar p$ scattering should very fast
decrease with the initial energy growth and seems to be very
small at LHC. As a consequence the dip in the differential cross
section clearly observed at LHC energy for $pp$ scattering should
manifest itself in $p\bar p$ scattering as well. It can be considered
as our prediction.

A detailed study of peculiar features that distinguish $pp$
and $p\bar p$ scattering needs a more careful analysis,
which is beyond the framework of this paper.

The authors are grateful to M.G.~Ryskin for helpful discussion.

This work has been supported by RSCF grant No 14 - 22 - 00281.


\begin{thebibliography}{**}

\bibitem{Dr} I.~M.~Dremin,
%``Elastic scattering of hadrons,''
Phys.\ Usp.\  {\bf 56} (2013) 3
[Usp.\ Fiz.\ Nauk {\bf 183} (2013) 3]
[arXiv:1206.5474 [hep-ph]].

\bibitem{RMK} M.~G.~Ryskin, A.~D.~Martin and V.~A.~Khoze,
%``Proton Opacity in the Light of LHC Diffractive Data,''
Eur.\ Phys.\ J.\ C {\bf 72} (2012) 1937

\bibitem{Khoze:2013dha}
V.~A.~Khoze, A.~D.~Martin and M.~G.~Ryskin,
%``Diffraction at the LHC,''
Eur.\ Phys.\ J.\ C {\bf 73} (2013) 2503
[arXiv:1306.2149 [hep-ph]].

\bibitem{Gotsman:2012rm}
E.~Gotsman, E.~Levin and U.~Maor,
%``Description of LHC data in a soft interaction model,''
Phys.\ Lett.\ B {\bf 716} (2012) 425
[arXiv:1208.0898 [hep-ph]].

\bibitem{Gotsman:2012rq}
E.~Gotsman, E.~Levin and U.~Maor,
%``Soft interaction model and the LHC data,''
Phys.\ Rev.\ D {\bf 85} (2012) 094007
[arXiv:1203.2419 [hep-ph]].

\bibitem{MerShab} C.~Merino and Y.~.M.~Shabelski,
%``Elastic pp Scattering at LHC Energies,''
JHEP {\bf 1205} (2012) 013
[arXiv:1204.0769 [hep-ph]].

\bibitem{Sel} O.~V.~Selyugin,
%``GPDs of the nucleons and elastic scattering at high energies,''
Eur.\ Phys.\ J.\ C {\bf 72} (2012) 2073
[arXiv:1201.4458 [hep-ph]].

\bibitem{LF} E.~M.~Levin and L.~L.~Frankfurt,
JETP Lett. {\bf 2} (1965) 65.

\bibitem{VH} J.~J.~J.~Kokkedee and L.~Van Hove,
%``Quark model and high-energy scattering,''
Nuovo Cim.\  {\bf 42} (1966) 711.

\bibitem{DDT}
Y.~L.~Dokshitzer, D.~Diakonov and S.~I.~Troian,
% ``Hard Processes in Quantum Chromodynamics,''
Phys.\ Rept.\  {\bf 58}, 269 (1980).

\bibitem{Shekhter} V.~M.~Shekhter, Yad.Fiz. {\bf 33} (1981) 817;
Sov. J. Nucl. Phys. {\bf 33} (1981) 426.

\bibitem{Avila}
R.~Avila, P.~Gauron and B.~Nicolescu,
%``How can the Odderon be detected at RHIC and LHC,''
Eur.\ Phys.\ J.\ C {\bf 49}, 581 (2007)
[hep-ph/0607089].

\bibitem{Glaub} R.~J.~Glauber. In "Lectures in Theoretical Physics",
Eds. W.~E.~Brittin etal., New York (1959), vol.1, p.315.

\bibitem{FG} V.~Franco and R.~J.~Glauber, Phys.Rev. {\bf 142}
(1966) 1195.

\bibitem{Bialas:1977xp}
A.~Bialas, K.~Fialkowski, W.~Slominski and M.~Zielinski,
%``Elastic p p Cross-Section and Multiple Scattering of Quarks,''
Acta Phys.\ Polon.\ B {\bf 8} (1977) 855.

\bibitem{Bialas:1982rd}
A.~Bialas and A.~Kolawa,
%``Wounded Nucleons in $\alpha^- \alpha$ Collisions at High-energies,''
Acta Phys.\ Polon.\ B {\bf 14} (1983) 539.

\bibitem{TOT1a} G. Antchev et al., TOTEM Collaboration, Europhys. Lett.
{\bf 96}, 21002 (2011).

\bibitem{TO1} TOTEM Collaboration, G. Antchev et al.,
% Measurement of proton-proton elastic scattering and total cross-section
% at S**(1/2) = 7-TeV,
Europhys.Lett. {\bf 101} (2013) 21002.

\bibitem{TO2} TOTEM Collaboration, G. Antchev et al.,
% Proton-proton elastic scattering at the LHC energy of s**(1/2) = 7 TeV,
Europhys.Lett. {\bf 95} (2011) 41001, [arXiv:1110.1385].

\bibitem{Gribov} V.N. Gribov, Sov.Phys.JETP {\bf 56} (1969) 892.

\bibitem{Froi} M. Froissart, Phys. Rev. {\bf 123} (1961) 1053.

\bibitem{Kar1} K.A. Ter-Martirosyan, Yad. Fiz. {\bf 10}, 1047 (1969).

\bibitem{Amos:1985wx}
N.~A.~Amos, M.~M.~Block, G.~J.~Bobbink, M.~Botje, D.~Favart,
C.~Leroy, F.~Linde and P.~Lipnik {\it et al.},
%``Measurement of Small Angle $\bar{p}p$ and Proton Proton Elastic
% Scattering at the CERN Intersecting Storage Rings,''
Nucl.\ Phys.\ B {\bf 262} (1985) 689.

\bibitem{Auger}
C. Auger {\it et al.}, UA4/2 Collaboration,
Phys.\ Lett.\ {\bf B316}, 448 (1993).

\bibitem{Abe:1993xx}
F.~Abe {\it et al.}  [CDF Collaboration],
%``Measurement of small angle $\bar{p}p$ elastic scattering at
% $\sqrt{s} = 546$ GeV and 1800 GeV,''
Phys.\ Rev.\ D {\bf 50} (1994) 5518.

\bibitem{Bozzo:1985th}
M.~Bozzo {\it et al.}  [UA4 Collaboration],
%``Elastic Scattering at the {CERN} {SPS} Collider Up to
% a Four Momentum Transfer of 1.55-{GeV}**2,''
Phys.\ Lett.\ B {\bf 155}, 197 (1985).

\bibitem{Kwa} N.~Kwak, E.~Lohrmann, E.~Nagy, M.~Regler,
W.~Schmidt-Parzefall, K.~R.~Schubert, K.~Winter and A.~Brandt {\it et al.},
% ``Experimental Results on Large Angle Elastic $p p$ Scattering at the CERN ISR,''
Phys.\ Lett.\ B {\bf 58} (1975) 233.

\bibitem{Am} U. Amaldi et al., Phys. Lett. {\bf B66}, 390 (1977).

\bibitem{Amos:1990fw}
N.~A.~Amos {\it et al.}  [E-710 Collaboration],
%``$\bar{p}p$ elastic scattering at $\sqrt{s}$ = 1.8-TeV
% from |t| = $0.034-GeV/c^{2}$ to $0.65-GeV/c^{2}$,''
Phys.\ Lett.\ B {\bf 247} (1990) 127.

\bibitem{Breakstone:1985pe}
A.~Breakstone, H.~B.~Crawley, G.~M.~Dallavalle, K.~Doroba, D.~Drijard, F.~Fabbri,
A.~Firestone and H.~G.~Fischer {\it et al.},
%``A Measurement of $\bar{p} p$ and $p p$ Elastic Scattering in the Dip Region
% at $\sqrt{s}=53$-{GeV},''
Phys.\ Rev.\ Lett.\  {\bf 54} (1985) 2180.

\bibitem{Erhan:1984mv}
S.~Erhan, A.~M.~Smith, L.~Meritet, M.~Reyrolle, F.~Vazeille, R.~Bonino, A.~Castellina
and M.~Medinnis {\it et al.},
%``Comparison of $\bar{p} p$ and $p p$ Elastic Scattering With
% $0.6-{\rm GeV}^ < t < 2.1-{\rm GeV}^2$
% at the {CERN} {ISR},''
Phys.\ Lett.\ B {\bf 152} (1985) 131.

\bibitem{Aamodt:2010dx}
K.~Aamodt {\it et al.}  [ALICE Collaboration],
%``Midrapidity antiproton-to-proton ratio in pp collisions
% at $\sqrt{s} = 0.9$ and $7$~TeV measured by the ALICE experiment,''
Phys.\ Rev.\ Lett.\  {\bf 105}, 072002 (2010).

\bibitem{Merino:2009zz}
C.~Merino, M.~M.~Ryzhinskiy and Y.~.M.~Shabelski,
%``Possible odderon effects in inclusive hadron production,''
Eur.\ Phys.\ J.\ C {\bf 62} (2009) 491.

\end{thebibliography}
\end{document}